\documentclass[twocolumn,english,aps,pra,longbibliography,superscriptaddress, floatfix]{revtex4-1}
\usepackage[T1]{fontenc}
\usepackage{inputenc}
\usepackage[dvipsnames]{xcolor}
\usepackage{amsmath,amsfonts,amsthm}
\usepackage{soul}

\usepackage[ruled,vlined]{algorithm2e}
\usepackage{booktabs}
\usepackage{graphicx}
\usepackage{amssymb}
 \usepackage[colorlinks=true,urlcolor=blue,citecolor=blue,linkcolor=blue]{hyperref}

\makeatother

\begin{document}

\title{Boosting Sparsity in Graph Decompositions with QAOA Sampling}

\author{George Pennington}
\affiliation{The Hartree Centre STFC, United Kingdom}

\author{Naeimeh Mohseni}
\affiliation{E.ON Digital Technology GmbH, Essen, Germany}

\author{Oscar Wallis}
\affiliation{The Hartree Centre STFC, United Kingdom}

\author{Francesca Schiavello}
\affiliation{The Hartree Centre STFC, United Kingdom}

\author{Stefano Mensa}
\affiliation{The Hartree Centre STFC, United Kingdom}

\author{Corey O'Meara}
\affiliation{E.ON Digital Technology GmbH, Essen, Germany}

\author{Giorgio  Cortiana}
\affiliation{E.ON Digital Technology GmbH, Essen, Germany}

\author{V\'ictor Valls}
\affiliation{IBM Quantum and IBM Research Europe -- Dublin, Ireland}

\begin{abstract}
We study the problem of decomposing a graph into a weighted sum of a small number of matchings, a task that arises in network resource allocation problems such as peer-to-peer energy exchange. Computing such decompositions is challenging for classical algorithms, even for small instances.
To address this problem, we propose E-FCFW, a hybrid quantum-classical algorithm based on the Fully-Corrective Frank-Wolfe (FCFW) algorithm that incorporates a matching-sampling subroutine. We design a QAOA version of this subroutine and benchmark it against classical approaches (random sampling and simulated annealing) on demand graphs derived from complete, bipartite, and heavy-hex topologies. The quantum subroutine is executed using the Qiskit Aer state-vector and MPS simulators and on IBM Kingston hardware (7--111 qubits). On complete and bipartite graphs with 6--10 nodes, E-FCFW with QAOA yields consistently sparser decompositions than the classical baselines, and even beats the best-known solution for one instance. On heavy-hex graphs with 50, 70 and 100 nodes, E-FCFW with QAOA outperforms the other methods in terms of approximation error, demonstrating performance on utility-scale quantum hardware. For the largest graphs (100 nodes) E-FCFW with QAOA performs much better when using MPS circuit simulation, compared to using quantum hardware. This indicates that at this scale, the performance is severely impacted by hardware noise.

\end{abstract}

\maketitle


\section{Introduction}
\label{sec:introduction}
Quantum optimisation aims to solve classically intractable optimisation problems by exploiting quantum mechanical phenomena \cite{abbas2024challenges}. While early demonstrations in finance, logistics, and energy systems show promise \cite{brandhofer2022benchmarking, doi:10.1126/science.abo6587,  blenninger2024q, mohseni2024competitive, bucher2024evaluating}, current hardware limits progress to small scales. This has led to increasing interest in hybrid quantum-classical approaches \cite{PhysRevApplied.20.034062, ELLINAS2024110835}, where quantum routines assist classical solvers on carefully chosen subproblems.

In this paper, we use quantum optimisation to solve a key subroutine within a heuristic algorithm for a graph decomposition problem. Specifically, we aim to decompose a weighted graph into as few matchings as possible (subgraphs where no two edges share a node). This problem generalises the Birkhoff decomposition problem from bipartite graphs to arbitrary graphs \cite{brualdi1982notes,durkovic2017birkhoff}. Minimising the number of matchings in such decompositions is NP-hard for bipartite graphs \cite{dufosse2016notes}, and even state-of-the-art classical algorithms struggle on instances with as few as 10 nodes \cite{koch2025quantumoptimizationbenchmarklibrary}.

\begin{figure}[t!]
\centering
\includegraphics[width=0.7\columnwidth]{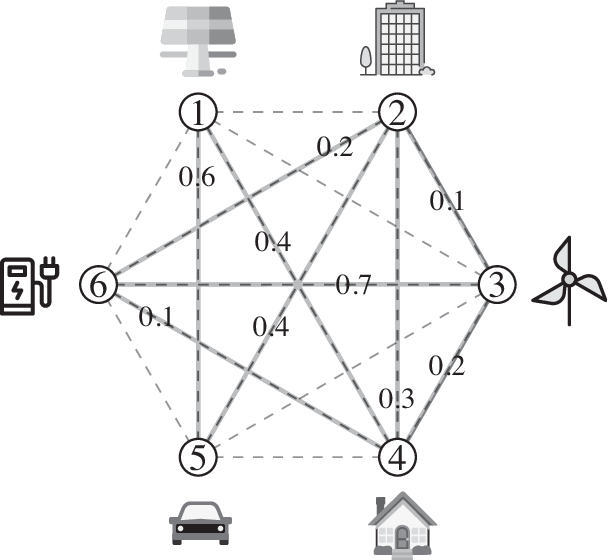}
\caption{Demand graph (thick edges) overlaid on a complete graph (dashed edges) with 6 nodes. Nodes represent devices (e.g., EVs), and each dashed edge indicates a feasible exchange of resources such as energy. The weight on each thick edge shows the fraction of time the corresponding devices wish to be connected. Fig.~\ref{fig:matchings_example} shows how the graph can be decomposed as the sum of four different matchings. }
\label{fig:graph_example}
\end{figure}

\begin{figure}
\centering
\includegraphics[width=0.8\columnwidth]{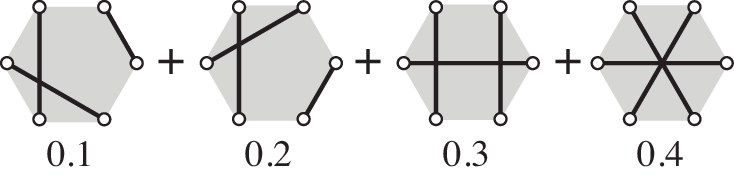}
\caption{Example of how the graph in Fig.~\ref{fig:graph_example} can be decomposed as the sum of graph matchings. A matching captures how nodes can exchange resources in practice. The number below each matching indicates the fraction of time each matching should be used to satisfy the demand graph in Fig.~\ref{fig:graph_example} }
\label{fig:matchings_example}
\end{figure}

This problem is not only of theoretical interest---better solutions are important for scheduling applications with operational constraints \cite{ghosh2024next,liu2015scheduling, promponas2023quantum}.  
As an example, Fig.~\ref{fig:graph_example} shows a power-grid scenario where nodes represent sites (e.g., EVs) and dashed edges indicate feasible energy-exchange links. The weight on each edge specifies the fraction of time two sites wish to exchange energy.  
The goal is to decompose this weighted graph into a small set of matchings, each representing a feasible configuration of simultaneous exchanges (Fig.~\ref{fig:matchings_example}). Fewer matchings mean that more of the desired exchanges can be scheduled within the available time.


\begin{figure}[t]
\centering
\includegraphics[width=0.9\columnwidth]{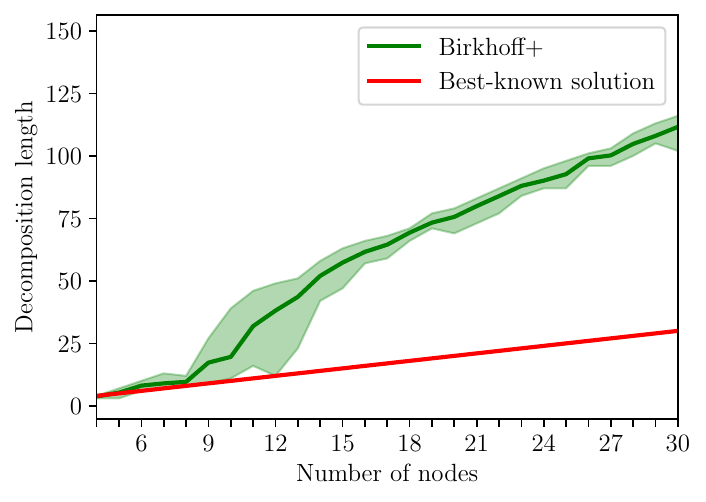}
\caption{Illustrating the decomposition length of the Birkhoff+ algorithm for bipartite graphs of size $n \in \{3,4,\dots,30\}$. There are 10 different graph instances for each graph size. Graphs are generated by sampling $n$ matchings and weights uniformly at random. The red line (best-known solution) indicates the number of matchings used to create the graph (i.e.~$n$). The light green area indicates the maximum and minimum decomposition length obtained for a graph size. The thick green line is the average decomposition length for all instances.  }
\label{fig:motivation_classic}
\end{figure}

Quantum techniques are appealing for this problem because graph decompositions rely on repeatedly computing matchings---an operation that can be naturally encoded as bitstrings and is therefore well suited to quantum optimisation---yet classical heuristics often fail to find matchings that lead to sparse decompositions. As shown in Fig.~\ref{fig:motivation_classic} for the Birkhoff+ algorithm \cite{valls2021birkhoff}, the number of matchings found is consistently higher than the best-known values, indicating a large gap between current solutions and what is actually achievable. 

This raises the central question investigated in this work: can quantum routines generate a more diverse set of high-quality matchings that lead to shorter graph decompositions? Intuitively, diversity in matchings reduces edge overlap, enabling shorter decompositions. Classical approaches typically focus on greedily finding a single maximum-weight matching, but this does not necessarily minimise the total number of matchings required for a full decomposition. Quantum algorithms, by contrast, explore the solution space differently and may yield a variety of high-weight matchings that result in shorter decompositions. Realising this potential requires algorithmic frameworks that can exploit multiple matchings and quantum subroutines that can compute matchings, even with current noisy hardware. To this end, we make the following contributions:

\begin{itemize}
\item \textbf{Graph decomposition problem.} 
We generalise the minimum Birkhoff decomposition problem from bipartite to arbitrary graphs (Sec.~\ref{sec:graph_scheduling}) and propose a mathematical model based on symmetric doubly substochastic matrices. The proposed model enables the modelling of low-degree networks that map efficiently to 50+ qubits on current hardware.

\item \textbf{E-FCFW algorithm + QAOA matching-sampling} 
We introduce E-FCFW, an extension of the Fully-Corrective Frank–Wolfe method that uses multiple matchings per iteration (Sec.~\ref{sec:algorithms}). We design a QAOA-based matching sampler that promotes edge diversity to improve decomposition quality (Sec.~\ref{sec:qaoa})

\item \textbf{Benchmarking.} We evaluate the performance of E-FCFW on different types of graphs---complete, bipartite, and heavy-hex---using different matching sampling techniques: random sampling, simulated annealing, and QAOA (Sec.~\ref{sec:experiments}). 
Our results show that E-FCFW + QAOA outperforms methods that use classical matching-sampling subroutines.

\end{itemize}

The rest of the paper is organised as follows. In Sec.~\ref{sec:graph_scheduling}, we describe the problem, present a mathematical model, and  discuss how the problem differs from the Birkhoff decomposition problem. In Sec.~\ref{sec:algorithms}, we introduce the new algorithm, E-FCFW, and explain how it utilises different subroutines for computing matchings, including QAOA. Sec.~\ref{sec:qaoa} shows how we can compute matchings with QAOA by solving a constrained discrete optimisation problem. Finally, Sec.~\ref{sec:experiments} presents the experimental evaluation and Sec.~\ref{sec:conclusions} concludes.

\section{Graph Decomposition Problem 
\\
\& Mathematical Model}
\label{sec:graph_scheduling}

In this section, we present a general and model-independent formulation of the graph decomposition problem for peer-to-peer energy exchange (Sec.~\ref{sec:model-independent}) and a specific mathematical formulation of the problem (Sec.~\ref{sec:model-dependent}). 

\subsection{The problem of decomposing an undirected graph as a small collection of matchings (model-independent formulation)}
\label{sec:model-independent}

Let $G(V,E)$ be an undirected graph with $V$ nodes and $E$ edges. An edge $e \in E$ connects two nodes $u, v \in V$. We use $e(u,v) = 1$ to denote that an edge $e$ connects nodes $u$ and $v$, otherwise $e(u,v) = 0$. Each edge $e \in E$ has a weight $w(e) \ge 0$ associated to it, where $w(e) = 0$ if there is no edge connecting nodes $u$ and $v$, i.e.~$e(u,v) = 0$.  

Let $\mathcal M(G)$ be a collection of subgraphs of $G$, where a subgraph $M \in \mathcal M(G)$ has  $V_M \subseteq V$ nodes and $E_M \subseteq E$ edges. We will call the graphs in $\mathcal M(G)$ matchings. For every matching $M \in \mathcal M(G)$, we have that $e_M(u,v) = 1$ if and only if $e_M(u,v') = 0$ and $e_M(u', v) = 0$ for all $u', v' \in V_M$ with $u' \ne u$ and $v' \ne v$. That is, if an edge $e_M \in E_M$ connects two nodes $u, v \in V_M$, these nodes are not connected to any other nodes. A matching graph does not need to be maximal (i.e.~contain the maximum possible number of edges) and not all nodes need to be connected to an edge. 

Let $\alpha_M \ge 0$ be a weight associated with every matching $M \in \mathcal M(G)$. The optimisation problem we want to solve is the following:
\begin{align*}
\begin{tabular}{lll}
$\underset{\alpha_M}{\text{minimise}}$ & $\displaystyle \sum_{M \in \mathcal M(G)} \alpha_M^0$ \\
subject to & $\displaystyle w(e) e(u,v)= \! \! \! \! \! \sum_{M \in \mathcal M(G)} \! \! \! \! \! \alpha_M e_M(u,v), \ \forall u,v \in V$
\end{tabular}
\end{align*}
where $0^0 = 0$. That is, we want to minimise the number of weights $\alpha_M$ that are non-zero subject to the constraint that $ w(e) e(u,v)= \sum_{M \in \mathcal M(G)} \alpha_M e_M(u,v)$. The variable $\alpha_M$ can be regarded as assigning the \emph{same} weights to all the edges in a subgraph $M$. Hence, the constraint captures that the sum of the subgraphs weights (masked by each edge) must be equal to the weight $w(e)$ in the original subgraph. 

We will always assume that the problem is feasible in the sense that for some given weights $w(e)$, there exists a collection of matchings and weights such that the constraints $w(e) e(u,v) = \sum_{M \in \mathcal M(G)} \alpha_M e_M(u,v)$ can be satisfied for all $u,v \in V$.


\subsection{Mathematical model \\ (model-dependent formulation)}
\label{sec:model-dependent}

In this section, we reformulate the problem described in Sec.~\ref{sec:graph_scheduling} into a mathematical model that is more suitable for algorithmic implementation. We can model the weighted graph in Sec.~\ref{sec:graph_scheduling} as  a \emph{symmetric} doubly substochastic matrix $D^* \in \mathbf R^{n \times n}$. That is, $D^*$ has non-negative entries and its rows and columns sum to less than one. Also, the matrix has zeroes in the diagonal since an edge on the graph must connect two different nodes. We use $D^*_{ij}$ to denote the $(i,j)$ entry of matrix $D^*$, which denotes the fraction of time that nodes $i$ and $j$ are willing to be connected, during which a commodity (e.g.~energy) can be exchanged. The assumption that $D^*$ is symmetric (i.e.~$D^*_{ij} = D^*_{ji}$) is for simplicity, and so our model does not consider the direction in which a commodity is exchanged. Similarly,  we can model a matching as \emph{symmetric} and \emph{binary} doubly substochastic matrix $P$, where the $(i,j)$ entry of $P$ represents how two nodes $i$ and $j$ in the original graph can exchange energy in practice. 

With the above mathematical model, we can cast the following optimisation problem:
\begin{align}
\begin{tabular}{ll}
$\underset{\alpha_l, P_l}{\textup{minimise}}$ & $\displaystyle \sum_{l=1}^k \alpha_l^0 $  \\
  subject to & $\displaystyle D^* = \sum_{l=1}^k \alpha_l P_l $, \quad  $P_l \in \mathcal P$\\
  &  $\displaystyle \alpha_l \ge 0, \ \displaystyle\sum_{l=1}^k \alpha_l \le 1$ 
\end{tabular}
\label{eq:main_opt_problem}
\end{align}
where $\mathcal P := \{ U \in \{0,1\}^{n \times n} \mid U^T \mathbf 1 \preceq \mathbf 1, U\mathbf 1 \preceq \mathbf 1, U_{ii} = 0 \ \forall \ i =1,\dots,n \}$ is the set of symmetric and binary doubly substochastic matrices---we use $\mathbf 1$ to denote the all-ones vector, and $x \preceq y$ indicates that vector $x$ is component-wise smaller than vector $y$. That is, the goal is to find a collection of symmetric and binary doubly substochastic matrices such that $D^* = \sum_{l=1}^k \alpha_l P_l$ for a given matrix $D^*$, where the weights must be non-negative and sum to less than or equal to one. The following is an example of the model for the graph in Fig.~\ref{fig:graph_example}:
\begin{align*}
\frac{1}{10}\left[
\begin{smallmatrix}
0 & 0 & 0 & 4 & 6 & 0 \\
0 & 0 & 1 & 3 & 4 & 2 \\
0 & 1 & 0 & 2 & 0 & 7 \\
4 & 3 & 2 & 0 & 0 & 1 \\
6 & 4 & 0 & 0 & 0 & 0 \\
0 & 2 & 7 & 1 & 0 & 0 
\end{smallmatrix}
\right]
& = 0.1
\left[
\begin{smallmatrix}
0 & 0 & 0 & 0 & 1 & 0 \\
0 & 0 & 1 & 0 & 0 & 0 \\
0 & 1 & 0 & 0 & 0 & 0 \\
0 & 0 & 0 & 0 & 0 & 1 \\
1 & 0 & 0 & 0 & 0 & 0 \\
0 & 0 & 0 & 1 & 0 & 0 
\end{smallmatrix}
\right]
+ 0.2 \left[
\begin{smallmatrix}
0 & 0 & 0 & 0 & 1 & 0 \\
0 & 0 & 0 & 0 & 0 & 1 \\
0 & 0 & 0 & 1 & 0 & 0 \\
0 & 0 & 1 & 0 & 0 & 0 \\
1 & 0 & 0 & 0 & 0 & 0 \\
0 & 1 & 0 & 0 & 0 & 0 
\end{smallmatrix}
\right] \\
& + 0.3
\left[
\begin{smallmatrix}
0 & 0 & 0 & 0 & 1 & 0 \\
0 & 0 & 0 & 1 & 0 & 0 \\
0 & 0 & 0 & 0 & 0 & 1 \\
0 & 1 & 0 & 0 & 0 & 0 \\
1 & 0 & 0 & 0 & 0 & 0 \\
0 & 0 & 1 & 0 & 0 & 0 
\end{smallmatrix}
\right]
+ 0.4 \left[
\begin{smallmatrix}
0 & 0 & 0 & 1 & 0 & 0 \\
0 & 0 & 0 & 0 & 1 & 0 \\
0 & 0 & 0 & 0 & 0 & 1 \\
1 & 0 & 0 & 0 & 0 & 0 \\
0 & 1 & 0 & 0 & 0 & 0 \\
0 & 0 & 1 & 0 & 0 & 0 
\end{smallmatrix}
\right]
\end{align*}

\subsection{Related models and classical algorithms}
\label{sec:background}
The problem and mathematical model presented are closely connected to the Birkhoff decomposition problem \cite{brualdi1982notes}. As mentioned in Sec.~\ref{sec:introduction}, our problem can be viewed as a generalisation of the Birkhoff decomposition problem, where we aim to decompose a general graph rather than a bipartite graph. The mathematical model we use is also related since we seek to decompose a doubly substochastic matrix as the weighted sum of symmetric and binary substochastic matrices. This can be viewed as a generalisation of the Birkhoff problem goal of decomposing a doubly stochastic matrix as the weighted sum of permutation matrices, reflecting the fact that, for general graphs, it may not always be possible for matchings to include every node.

The special characteristics of our problem (e.g.~symmetry) do not allow algorithms for decomposing doubly stochastic matrices to work, e.g.~\cite{liu2015scheduling,bojja2016costly,valls2021birkhoff}. Those algorithms rely on the fact the convex hull of permutation matrices is equal to the set of doubly stochastic matrices. However, that is not the case for symmetric doubly stochastic matrices since \emph{the convex hull of symmetric permutation matrices is only a \emph{subset} of the set of symmetric doubly stochastic matrices} \cite{katz1970extreme, cruse1975note}. In the next section, we will present a Frank-Wolfe (FW) algorithm, which is a more polyvalent type algorithm for decomposing an object (e.g.~a matrix) as the convex combination of different objects, which is often known as Carath\'edory's problem \cite{combettes2023revisiting}.

\section{Extended-FCFW}\label{sec:algorithms}

In this section, we present an extension of the Fully Corrective Frank-Wolfe (FCFW) algorithm tailored to our graph decomposition problem (Sec.~\ref{sec:efcfw}). The extension enables us to compute multiple matchings per iteration and compute the weights of the matchings such that the decomposition error is minimised, whilst providing a sparse decomposition. To provide context and build intuition, we begin by reviewing the classic FW algorithm (Sec.~\ref{sec:fw}), which establishes a connection to the optimisation problem in Eq.~\eqref{eq:main_opt_problem}. This foundation also motivates our choice to select maximum-weight matchings within the QAOA subroutine in Sec.~\ref{sec:qaoa}.

\subsection{FW background and motivation}
\label{sec:fw}

In short, FW is an algorithm devised for solving convex programs where the objective is a smooth convex function and the set is a polytope---such as the convex hull of a finite set of discrete elements.  In our graph decomposition problem, these discrete elements are symmetric binary substochastic matrices ($\mathcal P$) and $D^*$ is an element in their convex hull. The objective function we will use is
\begin{align}
 f(X) := \frac{1}{n^2} \left\| D^* - X\right\|_F^2,
 \label{eq:cost_function}
\end{align}
where $X$ is a matrix in the convex hull of symmetric binary substochastic matrices, and $\| \cdot \|_F$ denotes the Frobenius norm. 

The essence of the FW algorithm is to iteratively build solutions as convex combinations of the current approximation by selecting discrete elements from the feasible set. For example, let $\mathbb{M}$ denote a set of $k$ graph matchings. Then, we can write 
\begin{align*}
X_k :=\sum_{M \in \mathbb M} \alpha_M M
\end{align*} 
for some weights $\alpha_M \ge 0$, which can be selected so that $X_k$ is close to $D^*$. 

\begin{figure}[t!]
\centering
\includegraphics[width=0.5\columnwidth]{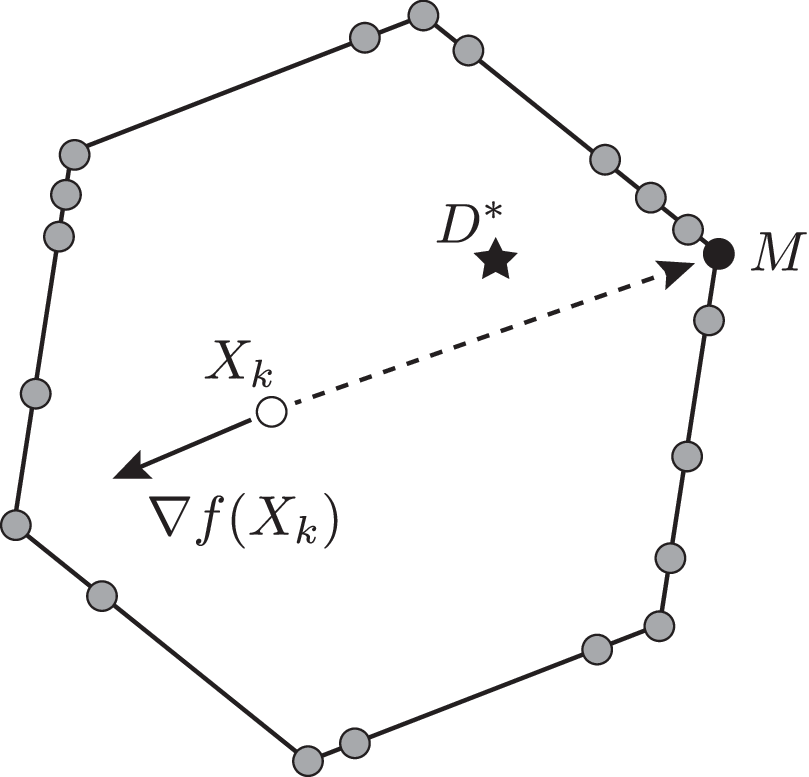}
\caption{Illustration showing one step of the Frank-Wolfe algorithm. The grey dots represent discrete objects to choose from (e.g.~matchings), and the polytope is the convex hull of the grey dots. The white dot ($X_k$) represents the current decomposition, and the black dot ($M$) is the maximum-weight matching selected by Frank-Wolfe with the update in Eq.~\eqref{eq:matching_selection}. The dashed arrow indicates the convex combination between $X_k$ and $M$, for which there exists a new graph decomposition ($X_{k+1})$ that is closer to the target demand matrix $D^*$ (black star).}
\label{fig:fw_intuition}
\end{figure}

FW selects how to add a new matching to set $\mathbb M$ by solving the following optimisation problem 
\begin{align}
\hat M \in \arg \min_{U \in \mathcal P}  \sum_{i,j=1}^n \nabla f(X)_{ij} U_{ij}.
\label{eq:matching_selection}
\end{align}
This update step is equivalent to finding a \emph{maximum-weight matching} on the graph defined by the matrix $(D^* - X_k)$---that is, finding the matching $\hat{M} \in \mathcal{P}$ that maximises the total weight of the selected edges in $(D^* - X_k)$. See Sec.~\ref{sec:model-dependent} where we map a graph to a matrix. The update can also be interpreted geometrically as finding a matching that aligns best with the negative direction of the gradient (see Fig.~\ref{fig:fw_intuition}). 

With update in Eq.~\eqref{eq:matching_selection}, Frank-Wolfe finds a collection of weights such that $X_k \to D^*$ as $k$ grows. The standard FW selects weights (also known as the learning rate) using a predefined rule. However, for our problem, and since we know the optimal solution $D^*$, it will be more convenient to recompute the weights with 
\begin{align*}
(\alpha_1,\dots,\alpha_{|\mathbb M|}) = \underset{u \in \Delta_{|\mathbb M|}}{\arg \min} \left \| D^* -  \sum_{M \in \mathbb M} u_M M ,\right\|_F^2
\end{align*}
where $\Delta_{|\mathbb M|}$ is the $|\mathbb M|$-dimensional simplex. This way of computing weights corresponds to the Fully-Corrective version of Frank-Wolfe \cite{lacoste2015global}.

\subsection{Extended-FCFW}
\label{sec:efcfw}

\begin{algorithm}[t]
	\caption{Extended Fully-Corrective FW (E-FCFW)}     \label{alg:FCFW}
	
         \KwIn{$n \times n$ demand matrix $D^*$; tolerance $\epsilon$}
         \SetKwInput{KwSet}{Set}
         \KwSet{\textup{Matchings set} $\mathbb{M} = \{\emptyset\}$, $k = 0$, $X_k = 0$, $\textup{error} = \frac{1}{n^2}\| D^*  - X_k \|_F^2$}
         
		\While{$\textup{error} > \epsilon$}{
        1) Find a collection of $d+1 \ge 1$ matchings $\{M_1,\dots, M_{d+1}\}$ in graph $D^* - X_k$ using the subroutine in Algorithm \ref{alg:matching-subroutine} and store them:
        \begin{align*}
             \mathbb{M} \leftarrow \mathbb{M} \cup \{M_1,\dots, M_{d+1}\}
        \end{align*}
        
        2) Compute weights by solving the following optimisation problem: 
        \begin{align*}
        \begin{tabular}{ll}
        $\underset{u_i \ge 0}{\textup{minimise}}$ & $\| D^* -\sum_{i=1}^{|\mathbb M|}{u_i M_i} \|_F^2$ \\
        $\text{subject to}$ & $\sum_{i=1}^{|\mathbb M|} u_i \leq 1 $\\
        & $\sum_{i=1}^{|\mathbb M|} u_i^0 = k +1 $\\
        & $M_i \in \mathbb M$
        \end{tabular}
        \end{align*}
        
        3) Update the weights and compute the new approximate decomposition and loss: 
         \begin{align*}
         \begin{tabular}{ll}
            (i)  & $\alpha \leftarrow \{u_j\} $\\
             (ii) & $X_{k+1}  \leftarrow  \sum_{M\in \mathbb M}{\alpha_M M}$ \\
             (iii) & $\textstyle \text{error}  \leftarrow \frac{1}{n^2}\| D^* - X_k \|_F^2$ \\
             (iv) & $k \leftarrow k+1$
         \end{tabular}
         \end{align*}
         }
         
    Remove any matchings with zero weight:
    \begin{align*}
    \begin{tabular}{ll}
        (i) & $\mathbb M \leftarrow  \{M_j \in \mathbb M : \alpha_{M_j} > 0 \}$ \\
        (ii)  & $\alpha \leftarrow \{u_j \in \alpha : u_j > 0 \} $\\
    \end{tabular}
    \end{align*}
		
    \KwOut{$\alpha$, $\mathbb M$}

\end{algorithm}

\begin{algorithm}[t]
\caption{Matching subroutine} \label{alg:matching-subroutine}

\SetKwInput{KwSet}{Set}
        \KwSet{$\mathbb{M}^\textup{FW} = \{\emptyset\}$ \textup{the first time Algorithm \ref{alg:FCFW} calls the subroutine to sample matchings.}}

    	\KwIn{Number of matchings to sample ($d$); Set $\mathbb M$; Approximate decomposition $X_k$; Demand matrix $D^*$.}

        1) Sample $d$ matchings from graph $D^* - X_k$ and store the matchings in set $\mathbb M^\textup{sample}$:
\begin{align*}
\mathbb M^\textup{sample} \leftarrow 
\{M_1, \dots, M_d\}
\end{align*}

        2) Find a maximum-weight matching $M^\textup{MW}$ in graph $D^* - \sum_{M \in \mathbb M^\textup{FW}} \alpha^\textup{FW}_M M$ where
\begin{align*}
(\alpha_1^\textup{FW},\dots,\alpha^\textup{FW}_{|\mathbb M^\textup{FW}|}) = \underset{u \in \Delta_{|\mathbb M^\textup{FW}|}}{\arg \min} \left \| D^* -  \sum_{M \in \mathbb M^\textup{FW}} u_M M \right\|_F^2
\end{align*}
Store the matching:
\begin{align*}
\mathbb M^\textup{FW} \leftarrow \mathbb M^\textup{FW} \cup M^\textup{MW}
\end{align*}  
	
	3) Pruning:
    
	\qquad a) $\mathbb M^\textup{FW} \leftarrow \textup{Unique}(\mathbb M^\textup{FW} )$
    
	\qquad b) $\mathbb M^\textup{sample} \leftarrow \textup{Unique}(\mathbb M^\textup{sample} ) \setminus\mathbb M^\textup{FW}  $


\KwOut{$(\mathbb M^\textup{sample} \cup \mathbb M^\textup{FW} ) \setminus \mathbb M$}

\end{algorithm}

Now, we present our extension of the FCFW algorithm (Algorithm \ref{alg:FCFW}). 
The algorithm takes as input the demand graph in matrix form ($D^*$) and a tolerance $\epsilon$ for which we want to find a decomposition such that $f(X) \le \epsilon$. The algorithm returns a collection of matchings and weights that approximate $D^*$. 
We initialise the initial approximation error to $\frac{1}{n^2} \| D^* \|_F^2$ and the set $\mathbb M$ to be the empty set. In each iteration $k\in \{0,1,2,\dots \}$ of the algorithm, we will store $d+1$ new matchings in set $\mathbb M$. Thus, at iteration $k$, $|\mathbb M| \leq (k+1)(d+1)$, however we limit the number of non-zero weights to $k+1$. The algorithm runs in loop until the desired approximation error $\epsilon$ is met. 

\subsubsection{E-FCFW Steps (Algorithm \ref{alg:FCFW})}
In the first step of the optimisation, Algorithm \ref{alg:matching-subroutine} is used to find a collection of $d+1$ matchings in the graph represented by the matrix $D^* - X_k$, with $d\in \{0,1,2,\dots \}$.  These matchings can be computed classically or with a quantum algorithm as we will discuss in more detail later. The matchings are stored in set $\mathbb M$.

The second step is to find a set of weights for the matchings in $\mathbb{M}$ that minimises the decomposition error, while using at most $k+1$ matchings, where $k$ is the current iteration count. Here, when we refer to a matching being ``used", we mean that it is assigned a non-zero weight. Limiting the number of matchings used to $k+1$ is important because it enables the algorithm to correct poor decisions made in earlier iterations. For example, suppose that at some iteration $k$, we achieve a decomposition with approximation error $\epsilon'$ using $k+1$ of the matchings in $\mathbb{M}$. Now suppose that by considering $\mathbb{M} \cup \{M_1,\dots,M_{d+1}\}$, we can attain the same (or better) approximation using a \textit{different} set of $k+1$ matchings. This means we can effectively discarded suboptimal matchings from the approximate decomposition, freeing the algorithm to select better ones in future iterations, which may lead to an even lower error. Moreover, by restricting the number of matchings per iteration, we ensure a fair comparison with the vanilla Fully Corrective Frank-Wolfe (FCFW) algorithm, for which $d = 0$ and only a single maximum-weight matching is sampled in each iteration.

The third step stores the weights, and computes the new approximate decomposition and the corresponding approximation error. 

Once the desired approximation error $\epsilon$ is reached, the final step is to return the matchings in $\mathbb{M}$ with non-zero weights, and their corresponding weights.

\subsubsection{Matching sampling subroutine (Algorithm \ref{alg:matching-subroutine})}
 The subroutine takes as input the number $d$ of matchings to sample, set $\mathbb M$, matrix $D^*$, and the approximate decomposition $X_k$. The subroutine initializes a set of matchings $\mathbb{M}^\textup{FW}$ the first time this is called by Algorithm \ref{alg:FCFW}. 
 
 The main part of the subroutine that computes matchings (Algorithm \ref{alg:matching-subroutine}) is split into 3 blocks.  The first block samples matchings from the graph represented by the matrix $D^* - X_k$. The sampling can be carried out with a classical or quantum algorithm, e.g.~with QAOA as we will explain in Sec.~\ref{sec:qaoa}, but also by sampling matching uniformly at random or with simulated annealing.
 The second block computes a maximum-weight matching in graph $D^* - \sum_{M \in \mathbb M^\textup{FW}} \alpha_M^\textup{FW} M$ where the weights minimise the decomposition error using only the matchings in set $\mathbb{M}^\textup{FW}$. The weights can be computed efficiently by solving a convex quadratic program, and the maximum-weight matching can be computed in polynomial time with Blossom's algorithm \cite{Edmonds_1965, vazirani1994theory}. This step ensures that the set of matchings at iteration $k$ of the E-FCFW algorithm always contains the corresponding set of matchings at iteration $k$ of the vanilla FCFW algorithm ($d=0$).
 The third and final step is to prune the sets to only keep unique elements. The pruning is carried out first by removing repeated elements from set $\mathbb{M}^\textup{FW}$ (if any) and then from set  $\mathbb{M}^\textup{sample}$. At the end of the pruning, the algorithm returns $(\mathbb M^\textup{sample} \cup \mathbb M^\textup{FW}) \setminus \mathbb M$, which is a collection of matchings that were not present in set $\mathbb M$.


\section{Quantum Subroutine for 
Computing Matchings}
\label{sec:qaoa}

In this section, we present a QAOA subroutine for sampling matchings. Sec.~\ref{sec:qaoa_motivation} motivates the use of QAOA in our problem. Sec.~\ref{sec:max-matching-formulation} presents the problem  of finding maximum-weight matchings as a discrete optimisation problem, which we use in the QAOA model in Sec.~\ref{sec:qaoa_model}.

\subsection{Motivation}
\label{sec:qaoa_motivation}

We employ QAOA, a heuristic quantum algorithm that leverages quantum fluctuations to explore the search space. Our motivation is inspired by the observation that, while QAOA at low circuit depths is typically regarded as limited in solution quality, this apparent drawback may become an advantage in our setting. Since our application benefits from diversity near optimality rather than exact optimality, the approximate solutions generated by low-depth QAOA are not only acceptable but particularly valuable. 

Fig.~\ref{fig:fw_intuition} illustrates the intuition behind the approach. The grey dots represent the valid matchings, the polytope is the convex hull of the matchings, the white dot ($X_k$) is the graph decomposition at time $k$, and the black star ($D^*$) is the matrix we want to decompose. In addition, the black dot here represents the maximum-weight matching selected by FW with the update in Eq.~\eqref{eq:matching_selection}. The red dashed line indicates the ground state evolution of QAOA: from the ground state of the initial Hamiltonian (black square) to the ground state of the cost Hamiltonian (black dot). The red coloured area represents the set from which bitstrings (matchings) are sampled with QAOA. The set of matchings from which QAOA samples may be different from other sampling techniques (blue area). The red and blue dots represent, respectively, the matchings that we would select with QAOA or with another technique (e.g., random sampling).

\begin{figure}[t!]
\centering
\includegraphics[width=0.5\columnwidth]{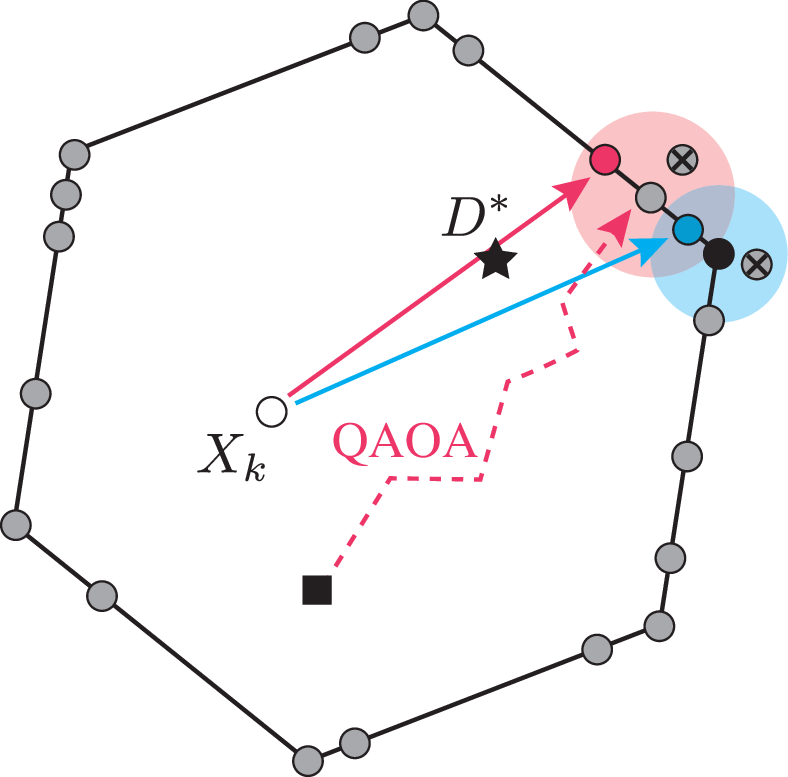}
\caption{Schematic illustration showing how QAOA may help to find better matchings explained in Sec.~\ref{sec:qaoa_motivation}. The grey dots with a cross (outside the polytope) indicate invalid matchings (i.e.~bitstrings) that may be selected during the sampling process. We note the figure is schematic as matchings are extreme points and therefore cannot be represented as the convex combination of two or more other matchings.}
\label{fig:qaoa_intuition}
\end{figure}

\subsection{QUBO Formulation for Maximum-Weight Matchings}
\label{sec:max-matching-formulation}

To apply QAOA to finding the maximally-weighted matchings of a
graph, we need a formulation that aligns with the underlying structure of the algorithm. Since QAOA can be interpreted as a parameterized quantum walk over the solution space, it is natural to encode edges as qubits rather than vertices. We formulate the problem as the following constrained binary optimisation model:
\begin{align}
\begin{tabular}{ll}
maximise & $\displaystyle \sum_{(u, v) \in E} \! \! \! \! w_{u v} x_{u v}$ \\
subject to & $\displaystyle \sum_{(u', v') \in \Gamma(u, v)} \! \! \! \! \! \! \! \! \! \! x_{u v} x_{u' v'} = 0, \quad  \hfill \forall (u,v) \in E$ \\
& $x_{u v} \in\{0,1\}, \quad \hfill \forall (u, v) \in E$
\end{tabular}
\label{eq:binary_model}
\end{align}
Here, $x_{u v}$  is a binary variable that takes the value $1$ if edge $(u,v)$ is included in the matching and zero otherwise. $w_{u,v}$ represents the weight of edge $(u,v)$ and $\Gamma (u,v)$ is the set of edges adjacent to edge $(u,v)$.
The constraint ensures that if an edge $(u,v)$ is included in the matching, all adjacent edges $(u',v') \in \Gamma(u,v)$ are not included.
With this encoding, every subgraph is represented as a unique bitstring, and every matching is a bitstring satisfying the constraint. For instance, in a graph with four edges, the bitstring $0101$ represents a subgraph that includes only the second and fourth edges.

We transform the constrained binary optimisation problem into an unconstrained problem by introducing a cost function, to be minimised, including the primary objective along with a penalty term representing the constraint:
\begin{align}
C(x)=-\sum_{(u, v) \in E}  \! \!  w_{uv} x_{u v} + \lambda  \!  \! \! \! \! \! \! \sum_{\substack{(u, v) \in E \\
(u', v') \in  \Gamma(u,v)}} \!  \! \! \! \! \! \!  x_{u v} x_{u' v'}
\label{eq:cost_func}
\end{align}
where $x \in \{0,1\}^{|E|}$. In this formulation, bitstrings violating the constraint in Eq.~\eqref{eq:binary_model} incur a cost penalty of $\lambda$. A crucial aspect of this method is the choice of the penalty term. It must be large enough to effectively enforce the constraints, ensuring that invalid matchings receive a significant penalty. However, if the penalty is too large, it can overshadow the primary objective, making optimisation difficult by biasing the search too heavily toward feasibility rather than optimality. Provided that $\lambda$ is chosen appropriately, the maximum-weight matching minimises Eq.~\eqref{eq:cost_func}.

Note that alternatively, we could incorporate constraints by embedding them directly into the mixing Hamiltonian, which restricts the search space to only feasible solutions by ensuring that quantum evolution remains within the valid subspace \cite{hadfield2019quantum}.
However, this method typically requires multi-qubit gates,
which are resource-intensive to decompose into two-qubit gates. In contrast, incorporating constraints into the penalty is more resource-efficient. 

\subsection{QAOA Model}
\label{sec:qaoa_model}

We use a standard QAOA model with one layer to implement the problem in Eq.~\eqref{eq:cost_func} 
\cite{farhi2014quantumapproximateoptimizationalgorithm}. The Hamiltonian that encodes the cost function is
\begin{align}
\hat{H}_C& =-\sum_{j \in E} w_j \left(\frac{1-\hat{Z}_j}{2}\right) \notag \\
 & \qquad +\lambda  \sum_{j \in E} \sum_{k \in  \Gamma(j)} \left(\frac{1-\hat{Z}_j}{2}\right) \left(\frac{1-\hat{Z}_k}{2}\right),
\label{eq:cost_hamiltonian}
\end{align}
where $\hat Z_j$ is the Pauli-Z operator acting on qubit $j$ and $w_j \ge 0 $ is the weight assigned to edge $j \in E$. Recall from Sec.~\ref{sec:max-matching-formulation} that we encode edges to qubits.
 The system is initially prepared in the state $|+\rangle^{\otimes N}$ where  $N:=|E|$ corresponds to the number of edges in the graph. Then, we apply evolution under the cost and mixing Hamiltonians $\hat U_C(\gamma)=\exp(-i \gamma \hat{H}_{\rm C})$ and $\hat U(\beta)=\exp(-i \beta \sum_{j=1}^N \hat{X}_{j})$ where $\hat X_j$ is the Pauli-X operator acting on qubit $j$, and $\gamma \in[0, 2\pi]$ and $\beta \in[0, \pi]$ are  variational parameters. This generates the following variational quantum state
\begin{align*}
|\psi( {\beta}, {\gamma})\rangle = \hat{U}_M(\beta)\hat{U}_C(\gamma)|+\rangle^{\otimes N}.
\end{align*}

The original cost function, Eq. \eqref{eq:cost_func}, is given by the expectation value of the cost Hamiltonian:
\begin{align*}
C({\gamma}, {\beta})=\left\langle\psi({\gamma}, {\beta})\right| H_C\left|\psi({\gamma}, {\beta})\right\rangle,
\end{align*}
where a classical computer is then used to search for the optimal parameters $
\left({\gamma}^*, {\beta}^*\right)=\arg \min _{{\gamma}, {\beta}} \ C({\gamma}, {\beta})$.
Once the optimal parameters $\left({\gamma}^*, {\beta}^*\right)$ are found, the variational circuit is executed multiple times with these fixed parameters to sample bitstrings $x \in\{0,1\}^{|E|}$ from the quantum state $\left|\psi\left({\gamma}^*, {\beta}^*\right)\right\rangle$. Each sampled bitstring corresponds to a candidate matching, and a few ones with the highest weight are selected as the final output.

\textbf{Circuit scaling.} The number of two-qubit gates in the QAOA ansatz increases rapidly with the number of edges in the graph. To illustrate this, consider a complete graph with $n$ nodes. In such a graph,  each node has degree $n-1$ and is adjacent to $2(n-2)$ other edges. Since our QAOA cost function, Eq.~\eqref{eq:cost_func}, requires the implementation of two-qubit \texttt{Rzz} gates between adjacent edges, this means the number of non-local gates in the QAOA ansatz, and in turn the circuit depth, can quickly become too large for current current devices. For example, in the 9-node graphs we will consider in Sec.~\ref{sec:random_fully_connected_graphs}, the two-qubit gate depth of the QAOA circuits ranges from 120 to 222 (after transpilation for IBM Sherbrooke), despite only requiring 20 to 25 qubits.

\subsection{Illustrative Example}

\begin{figure}[t]
    \centering
    \includegraphics[width=\linewidth]{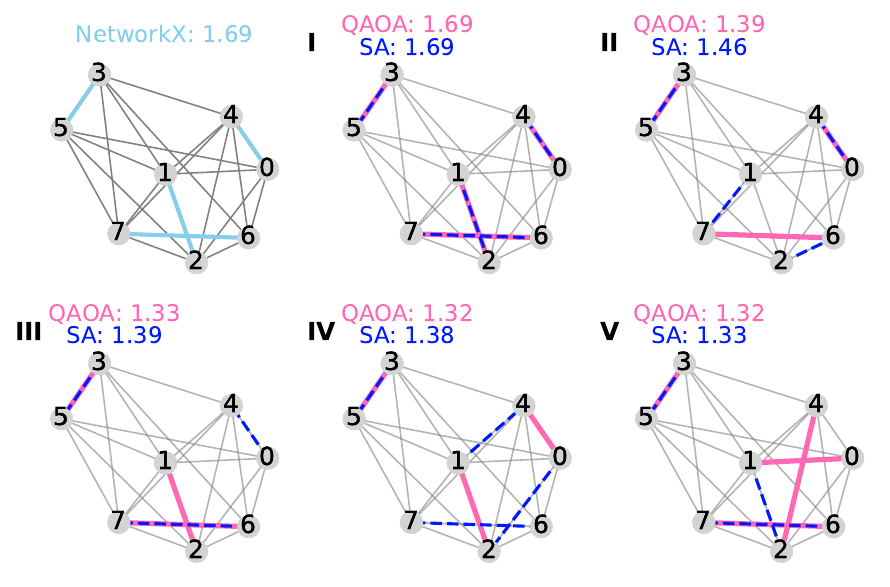}
    \caption{Maximum matching found by NetworkX for the graph with $n=8$ and $id=1$, along with the first 5 top matchings sampled using QAOA and SA. Each panel's title indicates the total weight of the corresponding matching. Recall that a maximum-weighted matching is a solution to the optimization in Eq.~\eqref{eq:matching_selection}.
    \label{fig:matchings}}
\end{figure}

Fig.~\ref{fig:matchings} shows the matchings sampled by QAOA for an 8-node random graph and compares it to the maximum-weight matching found by NetworkX \cite{hagberg2008exploring} and the matchings sampled with Simulated Annealing (SA). The figure shows the first five solutions sampled by QAOA and SA. Observe that the first sampled solution from both QAOA and SA (Panel I) matches the maximum-weight matching found by NetworkX. In the remaining sampled matchings, QAOA and SA often differ, with SA generally sampling matchings with higher total weight. Only the third solution sampled by SA coincides with the second-best solution found by QAOA.

\section{Benchmarking}
\label{sec:experiments}

In this section, we benchmark the E-FCFW algorithm across various graphs and matching sampling techniques. We begin by outlining the benchmarking methodology (Sec.~\ref{sec:methodology}), followed by a description of the experiments setup (Sec.~\ref{sec:setup}) and results (Sec.~\ref{sec:experiments_results}). All problem instances, source code, and performance results are available at \cite{qaoa-sparse-graph-repo}.

\subsection{Methodology: Goals, Limitations and Reporting}\label{sec:methodology}

The goal of our experiments is to determine whether sampling matchings using QAOA can help E-FCFW produce sparser decompositions compared to FCFW and E-FCFW with other matching sampling techniques. It is important to note that our findings are specific to the particular graphs and algorithm settings used in these experiments and may not generalise to other scenarios. In our case, the graph topologies we can consider is limited, as circuit depth increases rapidly with the number of edges (see discussion in Sec.~\ref{sec:qaoa_model}). Consequently, we restrict our experiments to small, densely connected graphs (e.g.~complete graphs) and larger, sparsely connected ones (e.g.~heavy-hex graphs). As for algorithm parameters (QAOA, simulated annealing, etc.), we have selected the best configurations available to us. In addition, it is important to keep in mind that these algorithms are stochastic, and repeated runs---even with identical settings---can yield different results, particularly in the quantum hardware experiments. 

In the following section, we discuss the performance of the algorithms in terms of the sparsity of the decomposition and the approximation error. However, other metrics such as runtime, resource usage, and the modeling approach are also of interest. To provide a comprehensive evaluation, we follow the quantum optimisation benchmarking guidelines outlined in \cite{koch2025quantumoptimizationbenchmarklibrary}. Table \ref{table:benchmarking_table} presents an example of the data we report for an experiment described in Sec.~\ref{sec:hardware_graphs}. 

\begin{table}[t]

\footnotesize{
\begin{center}
\begin{tabular*}{\linewidth}{l|l}

\toprule
\textbf{Problem Identifier} & Instance \#1 (Subgraph of IBM Sher-\\ & brooke with 50 nodes  and 53 edges)  \\
\textbf{Submitter}          & George Pennington, Oscar Wallis  \\
& Naeimeh Mohseni \\
\textbf{Date}               & July 18th, 2025 \\

\midrule
\textbf{Reference} & This paper\\

\midrule
\textbf{Best Objective Value} & 27 \\ 
\textbf{Optimality Bound}     & N/A \\

\midrule
\textbf{Modeling Approach}       & E-FCFW with QAOA sampling \\
& and CPLEX weights recomputation. \\
& The following applies to the $k$-th \\
& algorithm iteration, $k=0,1,2,\dots$    \\
\textbf{\#Decision Variables}     & 53 (QAOA) + $12 (k+1)$ (CPLEX) \\ 
\textbf{\#Binary Variables}       & 53 (QAOA) + $ 6(k+1)$ (CPLEX)\\ 
\textbf{\#Integer Variables}      &  N/A \\ 
\textbf{\#Continuous Variables}   & $6(k+1)$ (CPLEX)  \\ 
\textbf{\#Non-Zero Coefficients}  & 
$12 (k+1) + 2 $ (CPLEX)  \\ 
\textbf{Coefficients Type}        & integer \\
\textbf{Coefficients Range}       & $\{0,1,k+1\}$ (CPLEX) \\
      
\midrule 
\textbf{Workflow}          & Each iteration $k$ of E-FCFW calls: \\
                           & 1) QAOA sampling\\
                           & 2) CPLEX weight recomputation \\
\textbf{Algorithm Type}    & Stochastic \\ 
\textbf{\#Runs}            & 1 \\
\textbf{\#Feasible Runs}   & 1 \\
\textbf{\#Successful Runs} & 0 \\
\textbf{Success Threshold} & $10^{-6}$ \\
\midrule
\textbf{Hardware Specification}& QPU: \emph{ibm\_kingston} \\
\midrule
\textbf{Total Runtime}    & 761s \\
\textbf{CPU Runtime}      & 651s \\
\textbf{GPU Runtime}      & N/A \\
\textbf{QPU Runtime}      & 110s \\
\textbf{Other HW Runtime} & N/A \\

\bottomrule
\end{tabular*}
\end{center}
}
\label{table:benchmarking_table}
\caption{Performance metrics following the benchmarking guidelines in \cite{koch2025quantumoptimizationbenchmarklibrary}. The data in the table corresponds to the heavy-hex experiment with 50 nodes, instance \#1, and 53 qubits in Sec.~\ref{sec:hardware_graphs}.}
\end{table}

\subsection{Experiments setup}\label{sec:setup}

\subsubsection{Graph topology and demand matrices} 
\label{sec:graph_setup}
\textbf{Topology.} We consider three types of graph topologies: \textit{complete graphs} (Sec.~\ref{sec:random_fully_connected_graphs}), \textit{bipartite graphs} (Sec.~\ref{sec:random_bipartite_graphs}), and \textit{heavy-hex graphs} (Sec.~\ref{sec:hardware_graphs}). Complete and bipartite graphs have high-degree nodes, which makes the circuit depth of the QAOA ansatz grow quickly with the number of nodes. For this reason, we only consider complete and bipartite graphs of size 6-9 nodes (9-25 qubits) and 6, 8, 10 nodes (7-24 qubits) respectively, using state-vector circuit simulation. Heavy-hex graphs are subgraphs of the qubit connectivity of IBM Eagle and Heron processors and have low-degree nodes. For this type of graph, we will run experiments with graphs of size 50, 70 and 100 nodes, which require 52-111 qubits, using MPS circuit simulation and real quantum hardware execution.

\textbf{Demand matrices.} For all of our experiments, we use a consistent approach for constructing the demand matrices for a given graph topology. For a given underlying topology (complete, bipartite, heavy-hex), a demand matrix for a graph with $n$ nodes is constructed as follows:

\begin{enumerate}
\item Sample $n$ matchings in the underlying topology uniformly at random and map them to $n$ binary substochastic matrices $\{M_1,\dots,M_n\}$. The matchings may not necessarily be unique. This can be viewed as picking random matchings in a graph containing every possible allowed edge in the desired topology.

\item Select weights $\{\alpha_1,\dots,\alpha_n\}$ uniformly at random such that $0 \le \alpha_i \le 1$ for all $i=1,\dots,n$ and $\sum_{i=1}^n \alpha_i = 1$, and construct a convex combination with the selected permutation matrices, i.e.~$D^* = \sum_{i=1}^n \alpha_i M_i$. 
\end{enumerate}

This approach guarantees that $D^*$
  lies within the convex hull of the graph matchings and that a decomposition with at most $n$ matchings exists. We generate 10 demand matrices per graph topology. 

\subsubsection{Algorithms}\label{sec:general_algorithm_settings}

For all of our experiments, we use the following settings:

\begin{itemize}
    \item Decomposition approximation tolerance: $\epsilon=10^{-6}$.
    \item Number of matchings to sample in each iteration: $d=5$.
    \item Sampling methods: QAOA (Sec.~\ref{sec:qaoa}), random sampling, and simulated annealing. 
    \item Penalty term: $\lambda=0.2\sum_{j \in E} w_j$ (see Eq. \ref{eq:cost_func}).
\end{itemize}

Experiment-specific settings can be found in the corresponding sections.

The matching sampling subroutine in step 1) of Algorithm \ref{alg:matching-subroutine} is performed as follows. First, we sample $k$ bit-strings using either QAOA, random sampling, or simulated annealing. For random sampling, we assign the values 0 and 1 to each bit with equal probability. We then discard any bit-strings which do not correspond to valid matchings. For the remaining matchings, we calculate their weights and pick the $d=5$ with the highest weight. If there are fewer than $d=5$ valid matchings, we pick as many as possible. We use $k=10000$ for QAOA and random sampling, and $k=1000$ for simulated annealing (simulated annealing can be computationally intensive, and we observed no improvement when using $k=10000$ over $k=1000$). The maximum-weight matching in step 2) of Algorithm \ref{alg:matching-subroutine} is computed with NetworkX \cite{hagberg2008exploring}.

\subsection{Experiments Results}
\label{sec:experiments_results}

\begin{figure}[t]
    \centering\includegraphics[width=\linewidth]{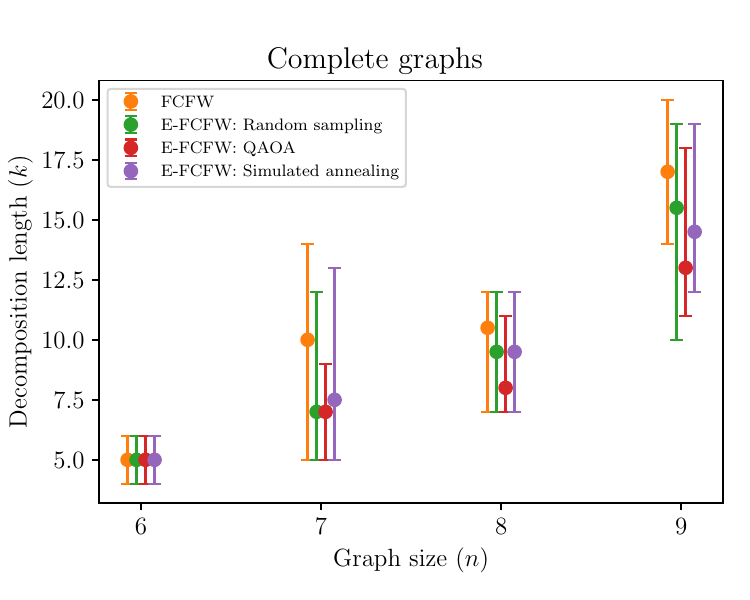}
    \caption{A plot showing the median decomposition length obtained by the FCFW and E-FCFW algorithms for fully-connected graphs of 6 to 9 nodes. 10 instances were used for each size. The error bars represent the maximum and minimum extent over the 10 instances. For the E-FCFW algorithm, we compare three different sampling methods for the matching subroutine (Algorithm \ref{alg:matching-subroutine}), namely QAOA, random sampling, and simulated annealing.}
    \label{fig:average_case_comparison_fully_connected}
\end{figure}

\subsubsection{Complete graphs}
\label{sec:random_fully_connected_graphs}

In this section, we evaluate the performance of the FCFW and E-FCFW algorithms for complete graphs with 6, 7, 8 and 9 nodes with demand graphs generated as described in Sec.~\ref{sec:graph_setup}. Recall there are 10 demand graphs for each topology. The number of edges with positive weights ranges from 9 to 25 for these instances. The algorithm settings are described in Sec.~\ref{sec:general_algorithm_settings}, where QAOA with one layer is performed classically using the Qiskit Aer state-vector simulator \cite{qiskit2024} (version 0.16.0). The hyperparameter optimisation is carried out with COBYLA (SciPy \cite{virtanen2020scipy} version 1.13.1) \cite{powell1994direct}.

\begin{figure}[t]
    \centering
        \includegraphics[width=\linewidth]{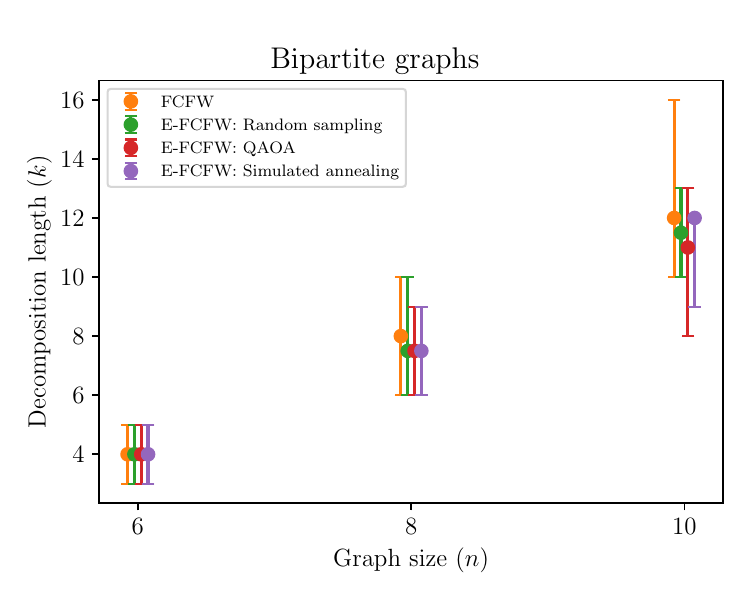}
    \caption{A plot showing the median decomposition length obtained by the FCFW and E-FCFW algorithms for bipartite graphs of 6, 8, and 10 nodes. 10 instances were used for each size. The error bars represent the maximum and minimum extent over the 10 instances. For the E-FCFW algorithm, we compare three different sampling methods for the matching subroutine (Algorithm \ref{alg:matching-subroutine}), namely QAOA, random sampling, and simulated annealing.}
    \label{fig:average_case_comparison_bipartite}
\end{figure}

\begin{figure*}[t]
    \centering
    \includegraphics[width=\linewidth]{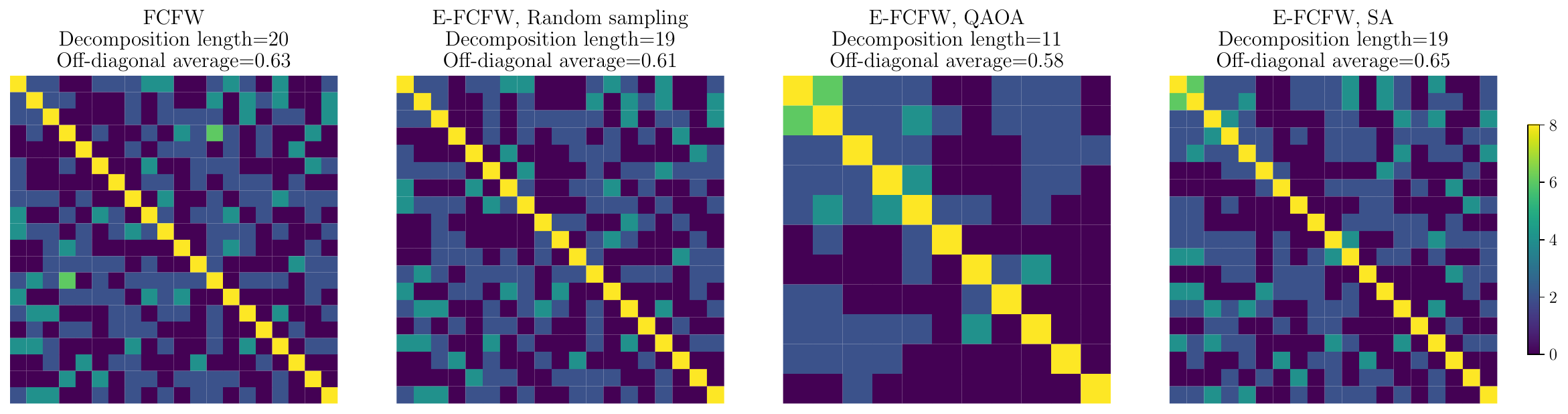}
    \caption{Matching overlap in the final decomposition for FCFW and E-FCFW using random sampling, QAOA, and simulated annealing on a complete graph with $9$ nodes (instance 2). The number of rows (and columns) reflects the number of matchings in the final decomposition length. Each cell in a row indicates the overlap of that matching with all others, measured by the number of shared edges. Diagonal entries represent the overlap of a matching with itself. }
    \label{fig:overlap}
\end{figure*}

\textbf{Results.} Our results are summarised in Fig.~ \ref{fig:average_case_comparison_fully_connected}, and 
Table \ref{tab:average_case_comparison_fully_connected} in Appendix \ref{sec:appendix_complete_and_bipartite} contains detailed results for all graph instances. Observe from the figure that the E-FCFW algorithm produces a lower median decomposition length than the FCFW algorithm for the 7, 8, and 9-node graphs, for all three sampling methods (QAOA, random sampling, and simulated annealing). For the 6-node graphs, the decomposition lengths reached by the FCFW and E-FCFW algorithms were identical for all 10 instances. This could be because at this small size, both algorithms found the minimum-length decomposition.
Our results demonstrate that the E-FCFW algorithm (Algorithm \ref{alg:FCFW}) tends to obtain sparser decompositions than the standard FCFW algorithm for the different sampling techniques. This result is expected from how we designed the sampling subroutine (Algorithm 
\ref{alg:matching-subroutine}). 

We find that applying QAOA to sample the matchings tends to result in lower median and average decomposition length than using random sampling or simulated annealing. However, the specific decomposition lengths vary greatly across instances. For example, with $n=9$ nodes, QAOA obtains the best results (not including ties) in 4 instances, simulated annealing in 1, and random sampling in 1. Also, we observe that while the median decomposition length is larger than the number of matchings used for generating the demand matrix ($n$), the decomposition length can be smaller than $n$ in some instances. 

\textbf{Discussion.} We hypothesize that the sampling subroutine based on low-depth QAOA yields shorter decomposition lengths than simulated annealing because the latter tends to find matchings with higher weight, which share many edges. Fig.~\ref{fig:overlap} shows the pairwise overlap between matchings (i.e.~the number of shared edges) for the instance with $n=9$, $id=2$. This instance exhibits a significant disparity in the decomposition lengths achieved by QAOA and simulated annealing (11 vs.~19). Observe from the figure that the matchings obtained by QAOA display lower overlap, indicating greater diversity.

 Of course, here we only consider the matchings obtained in the final decomposition, and the lower overlaps observed for lower decomposition lengths may simply be an inherent property when comparing different decompositions of the same graph. It remains an open area of investigation whether the E-FCFW algorithm can be improved by using a more sophisticated heuristic, one which selects matchings not just based on their weight, but tries to select matchings with high-weight and low-overlap with each other.

\subsubsection{Bipartite graphs}\label{sec:random_bipartite_graphs}

In this section, we consider the same setting as in Sec. \ref{sec:random_fully_connected_graphs} but for bipartite graphs. The graphs have a total of 6, 8, and 10 nodes (i.e.~bipartite graphs with 3, 4, and 5 nodes). Again, we use 10 demand matrices for each graph size, with the number of edges with positive weights ranging from 7 to 24. We use the algorithms as described in Sec. \ref{sec:general_algorithm_settings} with the same settings as in the experiments in Sec.~\ref{sec:random_fully_connected_graphs}.


\begin{figure*}[t]
    \centering
    \includegraphics[width=\linewidth]{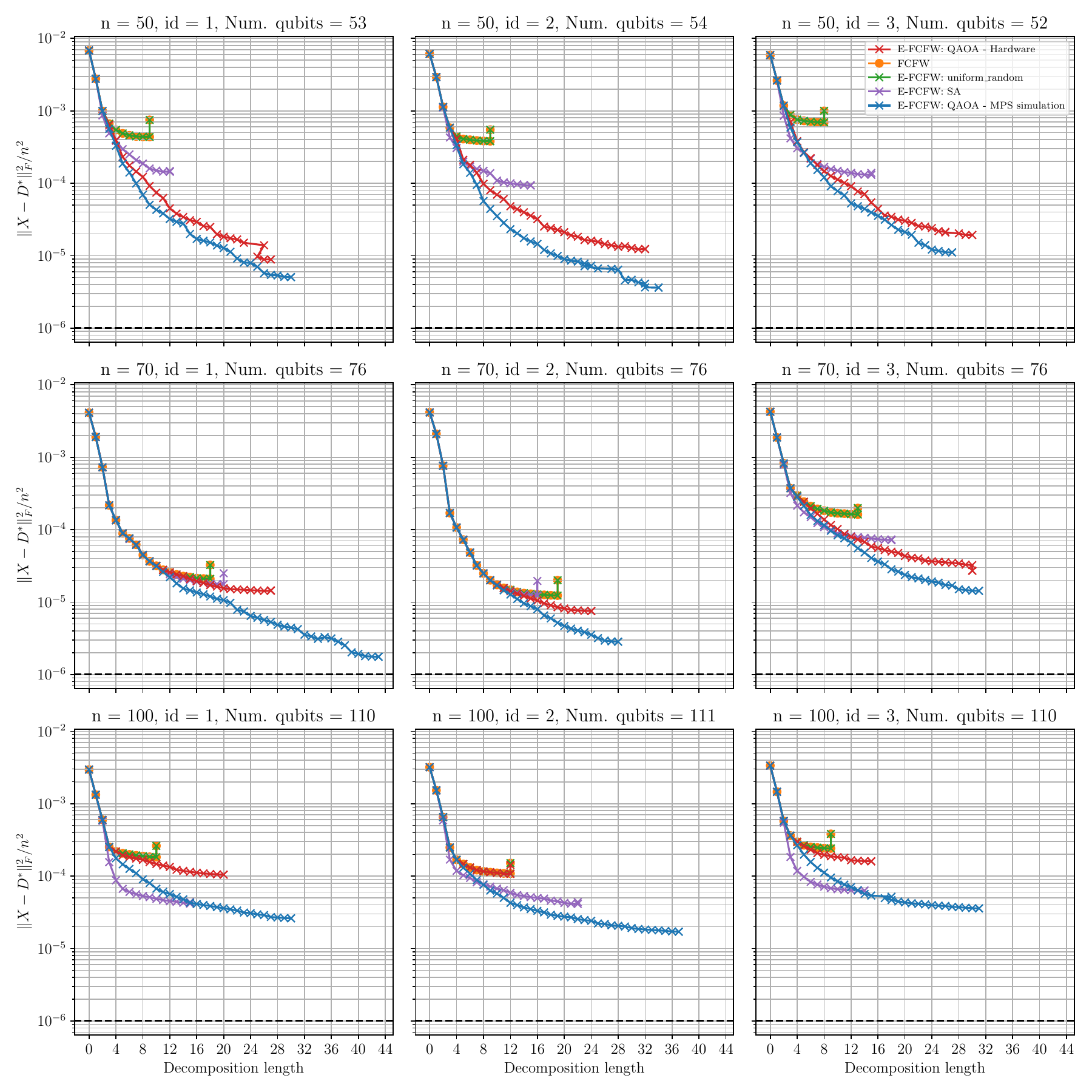}
        
    \caption{A plot showing the decomposition error vs decomposition length for nine heavy-hex topology graphs using the FCFW and E-FCFW algorithms. (Top) three 50-node graphs, with 53 (left), 54 (middle), and 52 (right) edges. (Middle) three 70-node graphs, all with 76 edges. (Bottom) three 100-node graphs, with 110 (left and right) and 111 (middle) edges. For the E-FCFW algorithm, we compare three different sampling methods for the matching subroutine (Algorithm \ref{alg:matching-subroutine}), namely QAOA, random sampling, and simulated annealing. For QAOA, we execute all circuits on IBM Kingston, and also compare with MPS circuit simulation. The black dashed line indicates the desired decomposition error $\epsilon = 10^{-6}$. The median two-qubit gate depths of the QAOA circuits were, from left to right, 38, 42.5, and 39.5 for the 50-node graphs, 50, 50, and 55 for the 70-node graphs, and 68.5, 69, and 72 for the 100-node graphs.}
    \label{fig:multi_panel_hardware_graphs}
\end{figure*}



\textbf{Results.} Our results are summarized in Fig.~ \ref{fig:average_case_comparison_bipartite}, and Table \ref{tab:average_case_comparison_fully_connected} in the Appendix contains detailed results for all instances. Observe from the figure that the results are similar to those of the previous experiment: the E-FCFW algorithm tends to obtain lower decomposition lengths than the FCFW algorithm, and the QAOA sampling is slightly better than the other sampling methods for $n=10$. Surprisingly, with $n=10$, the median obtained by random sampling is better than simulated annealing. In terms of performance per instance, we observe that the three matching sampling methods have a similar performance except in some instances. For example, for instance number 9 with 10 nodes, QAOA is able to obtain a decomposition length of 8 compared to 11 obtained by simulated annealing and random sampling. Such a result is remarkable because the decomposition length obtained is not just substantially smaller than the decomposition obtained with the other methods, but smaller than the number of matchings used to generate the demand matrix $D^*$ (see Sec.~\ref{sec:setup})---and the matchings used to generate the instance are all different in this case.

\textbf{Discussion.} The results for bipartite graphs are similar to the results we obtained for complete graphs. However, the major improvements we observe are when we compare the performance (decomposition length) per instance and not the median or average decomposition length. The result that random sampling performs better than simulated annealing suggests that sampling matchings with ``large weight'' might not always be the best strategy to obtain sparse decompositions. Recall that sampling matchings with large weights is motivated by the Frank-Wolfe update in Eq.~\eqref{eq:matching_selection}.

\subsubsection{Heavy-hex graphs}\label{sec:hardware_graphs}

In this section, we focus on a graph topology inspired by the heavy-hex architecture used in IBM quantum hardware. A key property of this topology is that each node has degree at most 3, which helps keep the number of non-local gates---and consequently the circuit depth---within practical limits for current quantum devices.

For our experiments, we evaluate performance on nine demand graphs constructed on the heavy-hex topology, following the methodology described in Sec.~\ref{sec:graph_setup}. We consider graphs with 50, 70, and 100 nodes, using three instances for each problem size. Their edge counts (which determine the required number of qubits) are 52--54 for the 50-node graphs, 76 for all 70-node graphs, and 110--111 for the 100-node graphs. 

We use the algorithms as described in Sec. \ref{sec:general_algorithm_settings}. However, in contrast with the previous experiments, we perform QAOA on both quantum hardware, using the IBM Kingston device, and with MPS circuit simulation with the Qiskit Aer matrix product state simulator \cite{qiskit2024} (version 0.16.0). For the results in this section, we fix the variational parameters to $(\beta,\gamma)=(-0.5,0.5)$. This simple parameter-setting strategy was inspired by the observation that the optimal parameters for 10 to 25 node heavy-hex graphs, trained using state-vector simulation, were consistently close to $(-0.5,0.5)$.

\textbf{Results.} Fig.~\ref{fig:multi_panel_hardware_graphs} shows the decomposition approximation error as a function of decomposition length for the various methods, for nine heavy-hex graphs. The decomposition length is usually equal to $k+1$, where $k$ is the iteration of the algorithm, but may be less than $k+1$ if some of the optimal weights are found to be equal to 0. In all plots, the curves for the FCFW algorithm and the E-FCFW algorithm with random-sampling lie on top of each other. This is because, at this scale, none of the randomly-sampled bitstrings corresponded to valid matchings, and hence they did not contribute to a sparser decomposition. This highlights the need for better sampling methods, such as QAOA and simulated annealing. 

For all of the graphs, we observe that none of the algorithms manage to reach a decomposition error lower than $\epsilon \le 10^{-6}$, as we had for the experiments in Sec.~\ref{sec:random_fully_connected_graphs} and Sec.~\ref{sec:random_bipartite_graphs}. 
However, the E-FCFW algorithm using QAOA and simulated annealing sampling manages to reach lower decomposition error than the FCFW algorithm. For all graphs, the decomposition error reached using QAOA sampling with MPS circuit simulation is lower than that reached using simulated annealing. This improvement is more significant for the 50 and 70-node graphs than for the 100-node graphs. For QAOA sampling using quantum hardware, we see the performance degrade as the problem size increases. For 50 nodes, E-FCFW with QAOA on quantum hardware outperforms simulated annealing, and performs similarly to QAOA using MPS simulation. For 70 nodes, the improvement of QAOA with quantum hardware over simulated annealing becomes less significant, and at 100-nodes, E-FCFW with QAOA on quantum hardware performs worse than simulated annealing. For the 100-node graphs, we observe an interesting feature: for instances 1 and 3, simulated annealing achieves a significantly better decomposition early in the algorithm, but the progress stalls as the number of matchings increases (i.e., for 16 matchings or more).  


\textbf{Discussion.} We observe that using MPS simulation for QAOA leads to lower decomposition error than using IBM Kingston, especially for the first 70-node graph. This suggests that hardware noise is affecting the performance of QAOA at this scale, and is likely why the performance of QAOA on quantum hardware for the 100-node graphs is poor. There are other factors which could affect the performance of QAOA at large scales, including: the parameter setting strategy losing its effectiveness when extrapolated too far from the graphs it was trained on, and the expressivity of the single-layer QAOA ansatz. However, since the main difference between the MPS simulation results and the hardware results is the presence of hardware noise, we suspect this is the dominant factor.

\begin{figure}[t]
    \centering
    \includegraphics[width=\linewidth]{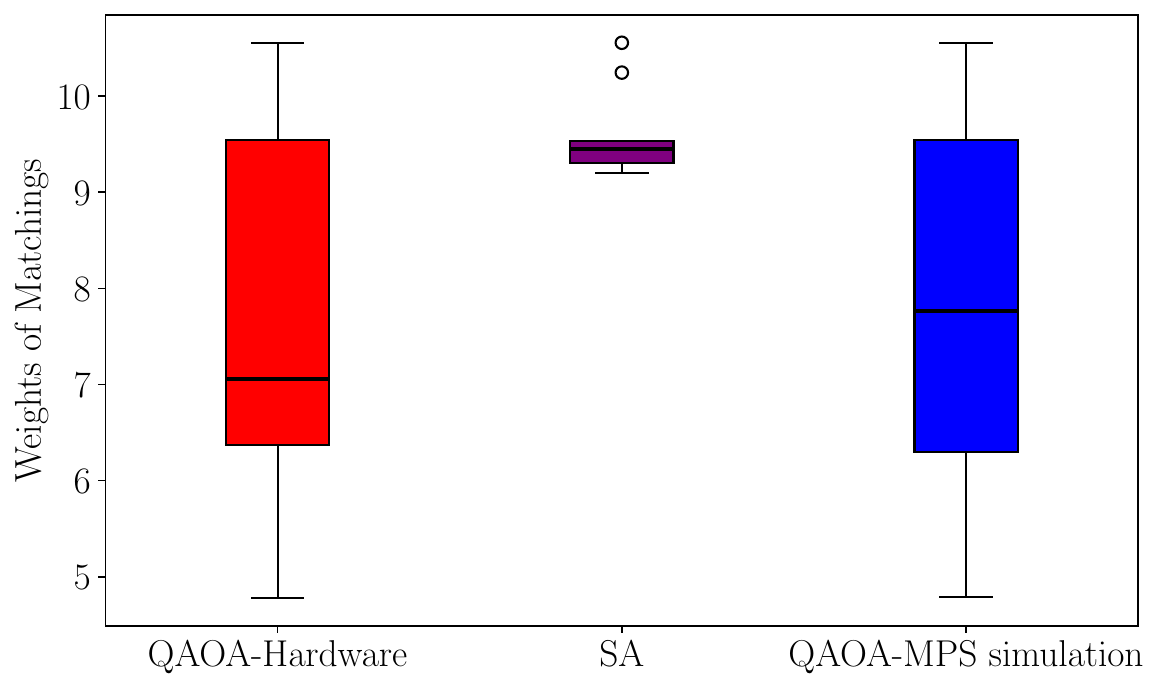}
    \caption{Distribution of the matchings weights in the final decomposition (instance $n=50$, $id=1$).
    \label{fig:boxplot}}
\end{figure}

Finally, Fig.~\ref{fig:boxplot} shows the distribution of matching weights in the final decomposition for the instance $n=50$ and $id=1$. As shown in the figure, QAOA (both on hardware and simulator) selects matchings with different total weights compared to SA. The above---together with the observations in Fig.~\ref{fig:overlap}---suggests that the E-FCFW algorithm benefits more from matching diversity rather than just weight.
Building on this observation, we hypothesize that low-depth QAOA can yield better decompositions than SA because it naturally generates a richer variety of matchings. In this way, our E-FCFW algorithm transforms what is typically considered a drawback of QAOA---its limited solution quality at low circuit depths---into a potential advantage. We emphasize, however, that this interpretation is based on our numerical observations rather than a formal proof and therefore requires further investigation.

\section{Conclusions}
\label{sec:conclusions}

This paper has studied the problem of decomposing a graph as the weighted sum of a small number of matchings. To address this problem, we have proposed a hybrid quantum-classical algorithm, E-FCFW, which extends the FCFW algorithm to allow it to use a matching-sampling subroutine. We showed how to design such a subroutine using QAOA to solve a constrained discrete optimisation problem that aims to find high-weight matchings in a graph. The motivation behind using QAOA is to find matchings with high-weight that use different edges in the graph.

We benchmark our approach on complete, bipartite, and heavy-hex graphs, conducting experiments with the Qiskit Aer state-vector simulator (9–25 qubits), the Qiskit Aer MPS simulator (52-111 qubits) and on IBM Kingston (52-111 qubits). Our results show that E-FCFW with QAOA consistently yields sparser decompositions (mean and median) than the other methods for small complete and bipartite graphs. For large heavy-hex graphs, E-FCFW with QAOA (executed using MPS simulation) outperforms the other methods in terms of approximation error.

For E-FCFW with QAOA executed on quantum hardware, we observe that the performance degrades with the problem size: outperforming the other methods for 50 and 70 nodes, but performing worse than simulated annealing for 100 nodes. We think this is likely due to hardware noise, but it could also be affected by our parameter setting strategy, and the single-layer QAOA ansatz used. Potential future research avenues to address this issue could include: using more sophisticated parameter setting strategies for QAOA, using deeper QAOA ans\"atze, and executing reduced-depth circuits on quantum hardware with higher qubit connectivity.

Another interesting research direction could be to explore different post-processing techniques for selecting the matchings in each iteration of the E-FCFW algorithm. For instance, instead of discarding bitstrings which violate the constraints, we could remove edges in the corresponding subgraphs until they correspond to valid matchings. Or, we could explore different selection criteria, rather than always selecting the highest-weight matchings. Such a selection criterion could be informed by overlaps between matchings, as discussed in Sec.~\ref{sec:random_fully_connected_graphs}.

\section{Acknowledgements}

This work was supported by the Hartree National Centre for Digital Innovation, a UK Government-funded collaboration between STFC and IBM. The authors would like to thank Stefan Woerner for his valuable feedback and suggestions.


\bibliographystyle{alpha}
\bibliography{references}


\clearpage
\appendix

\section{Complete and bipartite graph results}
\label{sec:appendix_complete_and_bipartite}

In this section we present the tabulated data used for Figs.~ \ref{fig:average_case_comparison_fully_connected} and \ref{fig:average_case_comparison_bipartite} in Sections \ref{sec:random_fully_connected_graphs} and \ref{sec:random_bipartite_graphs} respectively.

\begin{table}[h!]
\setlength{\tabcolsep}{1pt}  
\scriptsize
\caption{Tabulated results for Fig. \ref{fig:average_case_comparison_fully_connected}. The $n$ and \textit{id} columns specify the number of nodes in the graph and the graph instance respectively.}

    \centering
    \begin{tabular}{|c|c|c|c|c|c|c|c|c|c|}
    \hline
          &    & \multicolumn{4}{|c|}{Decomposition length} & \multicolumn{4}{|c|}{Decomposition approximation} \\ \hline
        &    & FCFW & \multicolumn{3}{|c|}{E-FCFW} & FCFW &  \multicolumn{3}{|c|}{E-FCFW} \\ \hline
        $n$ & id & MW & Rand & QAOA & SA & MW & Rand & QAOA & SA \\ \hline
        6 & 1 & 4 & 4 & 4 & 4 & 1.8E-10 & 5.6E-33 & 5.6E-33 & 5.6E-33 \\
        6 & 2 & 5 & 5 & 5 & 5 & 6.3E-10 & 6.3E-33 & 6.3E-33 & 9.3E-33 \\
        6 & 3 & 5 & 5 & 5 & 5 & 2.7E-10 & 1.1E-32 & 1.1E-32 & 1.1E-32 \\
        6 & 4 & 5 & 5 & 5 & 5 & 4.6E-10 & 6.2E-33 & 6.2E-33 & 7.2E-33 \\
        6 & 5 & 4 & 4 & 4 & 4 & 3.7E-10 & 1.7E-32 & 1.7E-32 & 1.7E-32 \\
        6 & 6 & 5 & 5 & 5 & 5 & 3.0E-10 & 4.7E-33 & 4.7E-33 & 6.5E-33 \\
        6 & 7 & 6 & 6 & 6 & 6 & 8.2E-12 & 4.1E-33 & 1.7E-33 & 2.9E-33 \\
        6 & 8 & 4 & 4 & 4 & 4 & 4.4E-07 & 4.3E-07 & 4.3E-07 & 4.3E-07 \\
        6 & 9 & 6 & 6 & 6 & 6 & 1.5E-10 & 8.3E-33 & 1.9E-32 & 2.8E-32 \\
        6 & 10 & 4 & 4 & 4 & 4 & 5.9E-10 & 4.8E-33 & 4.8E-33 & 4.5E-32 \\ \hline
        \multicolumn{2}{|c|}{{Average:}}& {\color{blue}{\textbf{4.8}}} & {\color{blue}{\textbf{4.8}}} & {\color{blue}{\textbf{4.8}}} & {\color{blue}{\textbf{4.8}}}    \\ \hline
        7 & 1 & 14 & 12 & 8 & 13 & 6.8E-10 & 6.2E-07 & 1.1E-32 & 6.2E-32 \\
        7 & 2 & 6 & 6 & 6 & 6 & 1.1E-09 & 3.8E-33 & 2.4E-32 & 1.2E-33 \\
        7 & 3 & 10 & 7 & 7 & 10 & 1.2E-07 & 1.5E-32 & 2.0E-33 & 3.8E-32 \\
        7 & 4 & 10 & 7 & 7 & 7 & 1.6E-10 & 5.0E-31 & 3.4E-32 & 2.5E-32 \\
        7 & 5 & 5 & 5 & 5 & 5 & 3.0E-10 & 4.3E-33 & 2.3E-32 & 4.9E-31 \\
        7 & 6 & 14 & 11 & 9 & 7 & 9.2E-08 & 9.1E-08 & 7.3E-33 & 9.0E-33 \\
        7 & 7 & 7 & 7 & 7 & 7 & 3.8E-10 & 4.0E-32 & 4.2E-32 & 1.8E-33 \\
        7 & 8 & 11 & 7 & 7 & 9 & 1.3E-09 & 1.8E-31 & 3.3E-32 & 1.3E-07 \\
        7 & 9 & 10 & 10 & 7 & 10 & 1.4E-09 & 6.7E-07 & 6.7E-32 & 3.7E-33 \\
        7 & 10 & 9 & 8 & 7 & 8 & 1.8E-10 & 4.5E-33 & 1.5E-31 & 9.3E-33 \\ \hline
        \multicolumn{2}{|c|}{{Average:}}& 9.6 & 8 & {\color{blue}{\textbf{7}}} & 8.2    \\ \hline
        8 & 1 & 11 & 9 & 10 & 9 & 2.0E-09 & 1.0E-33 & 5.4E-33 & 4.1E-33 \\
        8 & 2 & 12 & 12 & 9 & 12 & 6.9E-07 & 5.2E-07 & 8.0E-32 & 6.4E-07 \\
        8 & 3 & 10 & 10 & 8 & 10 & 3.1E-10 & 9.7E-33 & 1.3E-32 & 5.5E-33 \\
        8 & 4 & 11 & 11 & 11 & 11 & 8.8E-07 & 1.2E-32 & 2.6E-31 & 5.2E-33 \\
        8 & 5 & 11 & 10 & 9 & 11 & 1.4E-09 & 9.6E-07 & 1.8E-32 & 1.1E-32 \\
        8 & 6 & 11 & 10 & 8 & 11 & 3.0E-10 & 2.3E-34 & 5.6E-34 & 7.8E-33 \\
        8 & 7 & 8 & 8 & 7 & 8 & 2.4E-07 & 2.4E-07 & 2.5E-07 & 2.4E-07 \\
        8 & 8 & 8 & 8 & 8 & 8 & 2.1E-10 & 9.2E-34 & 8.1E-34 & 5.9E-33 \\
        8 & 9 & 7 & 7 & 7 & 7 & 7.4E-10 & 1.2E-33 & 1.1E-30 & 1.3E-33 \\
        8 & 10 & 8 & 8 & 8 & 8 & 8.8E-07 & 8.8E-07 & 1.6E-32 & 8.8E-07 \\ \hline
                \multicolumn{2}{|c|}{{Average:}}& 9.7 & 9.3 & {\color{blue}{\textbf{8.5}}} & 9.5    \\ \hline

        9 & 1 & 20 & 19 & 16 & 15 & 5.2E-07 & 4.8E-07 & 9.1E-07 & 5.1E-32 \\
        9 & 2 & 20 & 19 & 11 & 19 & 5.5E-07 & 5.4E-07 & 3.0E-07 & 8.5E-07 \\
        9 & 3 & 18 & 16 & 16 & 17 & 2.8E-07 & 2.8E-07 & 7.1E-07 & 2.6E-07 \\
        9 & 4 & 14 & 15 & 13 & 13 & 5.1E-07 & 4.0E-08 & 3.4E-07 & 5.5E-07 \\
        9 & 5 & 15 & 10 & 13 & 12 & 1.6E-09 & 4.0E-33 & 4.2E-07 & 1.7E-32 \\
        9 & 6 & 20 & 19 & 18 & 18 & 2.6E-07 & 8.4E-07 & 5.1E-10 & 2.7E-08 \\
        9 & 7 & 15 & 13 & 13 & 13 & 5.6E-07 & 7.4E-07 & 6.5E-07 & 7.4E-07 \\
        9 & 8 & 15 & 13 & 11 & 14 & 8.7E-10 & 6.2E-10 & 9.9E-07 & 1.7E-32 \\
        9 & 9 & 16 & 15 & 12 & 14 & 6.2E-07 & 8.5E-07 & 8.8E-07 & 5.4E-07 \\
        9 & 10 & 19 & 18 & 13 & 16 & 6.4E-07 & 8.1E-07 & 9.4E-07 & 6.1E-07 \\ \hline
                \multicolumn{2}{|c|}{{Average:}}& 17.2 & 15.7 & {\color{blue}{\textbf{13.6}}} & 15.1    \\ \hline

    \end{tabular}\label{tab:average_case_comparison_fully_connected}
\end{table}
\vfill

\text{}
\pagebreak 

\begin{table}
\setlength{\tabcolsep}{1pt}  
\scriptsize
\caption{Tabulated results for Fig. \ref{fig:average_case_comparison_bipartite}. The entries marked in red did not reach the target approximation error $\epsilon = 10^{-6}$.}
    \centering
    \begin{tabular}{|c|c|c|c|c|c|c|c|c|c|}
    \hline
          &    & \multicolumn{4}{|c|}{Decomposition length} & \multicolumn{4}{|c|}{Decomposition approximation} \\ \hline
        &    & FCFW & \multicolumn{3}{|c|}{E-FCFW} & FCFW &  \multicolumn{3}{|c|}{E-FCFW} \\ \hline
        $n$ & id & MW & Rand & QAOA & SA & MW & Random & QAOA & SA \\ \hline
        6 & 1 & 4 & 4 & 4 & 4 & 5.3E-10 & 2.2E-33 & 2.2E-33 & 2.2E-33 \\
        6 & 2 & 3 & 3 & 3 & 3 & 1.5E-10 & 1.6E-31 & 1.6E-31 & 1.6E-31 \\
        6 & 3 & 5 & 5 & 5 & 5 & 4.6E-10 & 8.1E-33 & 6.6E-33 & 8.1E-33 \\
        6 & 4 & 3 & 3 & 3 & 3 & 1.8E-10 & 2.5E-32 & 2.5E-32 & 2.5E-32 \\
        6 & 5 & 4 & 4 & 4 & 4 & 3.8E-10 & 1.7E-32 & 1.7E-32 & 1.7E-32 \\
        6 & 6 & 4 & 4 & 4 & 4 & 3.3E-10 & 2.0E-33 & 2.0E-33 & 2.0E-33 \\
        6 & 7 & 4 & 4 & 4 & 4 & 1.7E-10 & 3.0E-33 & 3.0E-33 & 3.0E-33 \\
        6 & 8 & 4 & 4 & 4 & 4 & 1.7E-10 & 3.6E-33 & 3.6E-33 & 9.0E-34 \\
        6 & 9 & 5 & 5 & 5 & 5 & 1.2E-10 & 9.4E-33 & 5.5E-33 & 4.0E-32 \\
        6 & 10 & 4 & 4 & 4 & 4 & 6.3E-10 & 4.6E-33 & 3.0E-33 & 3.0E-33 \\ \hline
                \multicolumn{2}{|c|}{{Average:}}& {\color{blue}{\textbf{4}}} & {\color{blue}{\textbf{4}}} & {\color{blue}{\textbf{4}}} & {\color{blue}{\textbf{4}}}    \\ \hline
        8 & 1 & 8 & 8 & 7 & 6 & 6.3E-10 & 6.5E-31 & 8.4E-07 & 2.4E-33 \\
        8 & 2 & 8 & 8 & 8 & 8 & 1.1E-07 & 1.1E-07 & 1.1E-07 & 1.1E-07 \\
        8 & 3 & 6 & 6 & 6 & 6 & 4.2E-10 & 2.0E-33 & 3.6E-32 & 1.7E-32 \\
        8 & 4 & 9 & 7 & 8 & 9 & 2.7E-10 & 2.4E-32 & 4.4E-33 & 8.1E-33 \\
        8 & 5 & 8 & 8 & 8 & 8 & 6.8E-10 & 3.1E-31 & 2.0E-32 & 2.0E-32 \\
        8 & 6 & 6 & 6 & 6 & 6 & 1.2E-10 & 1.2E-32 & 1.3E-30 & 1.7E-33 \\
        8 & 7 & 6 & 6 & 6 & 6 & 3.8E-10 & 3.3E-33 & 4.8E-33 & 8.9E-33 \\
        8 & 8 & 9 & 9 & 9 & 9 & 3.0E-08 & 2.9E-08 & 2.9E-08 & 2.9E-08 \\
        8 & 9 & 10 & 10 & 8 & 8 & 1.7E-09 & 1.6E-33 & 3.1E-31 & 7.0E-31 \\
        8 & 10 & 7 & 6 & 6 & 7 & 4.5E-10 & 6.5E-33 & 5.9E-32 & 2.8E-30 \\ \hline
                \multicolumn{2}{|c|}{{Average:}}& 7.7 & 7.4 & {\color{blue}{\textbf{7.2}}} & 7.3    \\ \hline
        10 & 1 & 16 & 13 & 13 & 12 & 1.8E-07 & {\color{red}{1.4E-06}} & 2.1E-07 & 9.4E-07 \\
        10 & 2 & 10 & 10 & 9 & 9 & 6.4E-07 & 6.4E-07 & 3.1E-07 & {\color{red}{1.2E-06}} \\
        10 & 3 & 10 & 10 & 8 & 10 & 1.7E-09 & 3.2E-31 & 3.6E-07 & 1.1E-31 \\
        10 & 4 & 12 & 12 & 12 & 12 & 1.3E-08 & 1.2E-08 & 1.2E-08 & 1.2E-08 \\
        10 & 5 & 12 & 11 & 11 & 12 & 8.2E-08 & 7.7E-07 & 1.1E-07 & 2.0E-10 \\
        10 & 6 & 13 & 13 & 11 & 12 & 7.8E-07 & 7.7E-07 & 2.2E-30 & 1.6E-07 \\
        10 & 7 & 13 & 13 & 12 & 12 & 5.1E-07 & 8.8E-07 & 8.3E-07 & 8.3E-07 \\
        10 & 8 & 10 & 10 & 10 & 10 & 6.2E-07 & 5.5E-07 & 8.5E-33 & 6.2E-07 \\
        10 & 9 & 12 & 11 & 8 & 11 & 5.1E-10 & 7.7E-33 & 8.7E-07 & 9.5E-33 \\
        10 & 10 & 12 & 12 & 12 & 12 & 6.2E-07 & 6.2E-07 & 1.3E-08 & 6.2E-07 \\ \hline
                \multicolumn{2}{|c|}{{Average:}}& 12 & 11.5 & {\color{blue}{\textbf{10.6}}} & 11.2    \\ \hline
    \end{tabular}
\label{tab:average_case_comparison_bipartite}
\end{table}

\end{document}